\documentclass[aps,prd,twocolumn,eqsecnum,showpacs,amsmath]{revtex4}
\usepackage[dvips]{color,graphicx,lscape}
\usepackage{amsfonts,amssymb,theorem}
\newcommand{\D}{{\rm d}}

{\theorembodyfont{\upshape}
}
{\theorembodyfont{\upshape}
}
{\theorembodyfont{\upshape}
}
{\theorembodyfont{\upshape}
}
{\theorembodyfont{\upshape}
}
{\theorembodyfont{\upshape}
}

\newcommand{\dalm}{\kern1pt\vbox{\hrule height 0.9pt\hbox{\vrule width
0.9pt\hskip 2.5pt\vbox{\vskip 5.5pt}\hskip 3pt\vrule width 0.3pt}\hrule height
0.3pt}\kern1pt}

\def\b2hat{ {\hat b}_2 }

\begin{document}

\title{
Decreasing entropy of dynamical black holes in critical gravity
}

\author{Hideki Maeda${}^a$}
\email{h-maeda@hgu.jp}
\author{Robert \v{S}varc${}^b$}
\email{robert.svarc@mff.cuni.cz}
\author{Ji\v{r}\'{\i} Podolsk\'y${}^b$}
\email{podolsky@mbox.troja.mff.cuni.cz}


\affiliation{
	${}^a$ Department of Electronics and Information Engineering, Hokkai-Gakuen University, Sapporo 062-8605, Japan. \\
	${}^b$ Institute of Theoretical Physics, Charles University, Prague, Faculty of Mathematics and Physics, V~Hole\v{s}ovi\v{c}k\'ach 2, 18000 Prague 8, Czech Republic.
}

\date{\today}

\begin{abstract}
Critical gravity is a quadratic curvature gravity in four dimensions which is ghost-free around the AdS background.
Constructing a Vaidya-type exact solution, we show that the area of a black hole defined by a future outer trapping horizon can shrink by injecting a charged null fluid with positive energy density, so that a black hole is no more a one-way membrane even under the null energy condition.
In addition, the solution shows that the Wald-Kodama dynamical entropy of a black hole is negative and can decrease.
These properties expose the pathological aspects of critical gravity at the non-perturbative level.
\end{abstract}


\maketitle


\section{Introduction}
It is known that the Minkowski vacuum in general relativity is dynamically stable against linear perturbations but the theory is not renormalizable, which is the main reason preventing perturbative quantization of gravity.
However, if the quantum theory of gravity exists, it is natural to expect that there appear higher-curvature terms as corrections to general relativity in the action of its low-energy classical theory and the resulting field equations contain higher-derivative terms in general.
In four dimensions, the coupling constants for the quadratic curvature terms are dimensionless in the units ${c=\hbar=1}$ and such terms are dominant in the high-energy scale.
For this reason, gravitation theories including quadratic curvature terms may be renormalizable~\cite{Weinberg:1974tw,Deser:1975nv,Stelle:1976gc}.
(See~\cite{Salvio2018} for a recent review on quadratic gravity.)

However, in addition to renormalizability, unitarity should be required for classical theories to be quantized in a perturbative manner.
Quadratic curvature theories admit two distinct maximally symmetric vacua in general, and usually there appear ghosts around both of them, which means that such theories are non-unitarity.
In this context, Einstein-Gauss-Bonnet gravity is a well-known exceptional case, of which action consists of a special quadratic combination of the curvature tensors and the field equations do not contain higher-derivative terms.
This theory is ghost-free around one of the maximally symmetric vacua which admits the general-relativistic limit~\cite{Zwiebach:1985uq,bb1996}.
However, this is the case only in higher dimensions because the Gauss-Bonnet term in the Lagrangian density becomes topological and the theory reduces to general relativity in four and lower dimensions.

Indeed, there exists a quadratic curvature gravity in three dimensions possessing similar properties to Einstein-Gauss-Bonnet gravity, which is now so-called BHT massive gravity~\cite{bht2009}.
Unlike Einstein-Gauss-Bonnet gravity, the field equations in this theory contain higher-derivative terms.
Nevertheless, it is ghost-free around the flat background~\cite{bht2009,bht2009b,no2009,ohta2012} and also around the anti-de~Sitter (AdS) background with a fine-tuning between the coupling constants~\cite{ls2009}.
Unfortunately, contrary to the earlier claim~\cite{oda2009}, this theory was found to be non-renormalizable~\cite{mo2012}.

Inspired by these results in three dimensions, a unitary quadratic curvature theory of gravity has been constructed in four dimensions, which is called critical gravity~\cite{criticalgravity}.
In spite that the field equations contain higher-derivative terms, similar to BHT massive gravity, it is ghost-free around the AdS background~\cite{criticalgravity,ohta2012}.
Higher-dimensional generalization of this critical gravity has been also achieved~\cite{criticalgravity-higher}, in which the maximally symmetric vacua are not necessarily unique due to the contribution of the Gauss-Bonnet term in the action.
Up to now, it is still not clear whether critical gravity and its higher-dimensional counterpart are renormalizable or not.

Although the maximally symmetric vacuum is not Minkowski but AdS in critical gravity, this is a remarkable classical theory holding unitarity.
Then a natural question arises: Is the theory well-behaving also in the spacetime much different from the maximally symmetric one?
This paper provides a negative answer to this question by constructing an exact solution representing an evolving black hole, defined by a future outer trapping horizon.
More concretely, the solution represents a shrinking black hole by the injection of matter fields satisfying the null energy condition, which is never realized in general relativity.
This shows that a black hole does not capture a concept of a one-way membrane under the energy condition in critical gravity.
Furthermore, our solution shows that the entropy of dynamical black holes can decrease under the null energy condition.
These properties exposes that the theory is pathological at the non-linear level.

The outline of the present paper is as follows.
In Sec.~\ref{sec2}, we summarize the field equations in the most general quadratic curvature gravity in arbitrary $n(\ge 4)$ dimensions and identify the parameter space for critical gravity.
A definition of a dynamical black hole in terms of the trapping horizon is also explained.
In Sec.~\ref{sec3}, we present several new Vaidya-type exact solutions for a null dust fluid together with a Maxwell field in the most general quadratic curvature gravity in four dimensions, which are compared with the corresponding solutions in general relativity and Einstein-Gauss-Bonnet gravity in arbitrary dimensions.
In Sec.~\ref{sec4}, we study the properties of dynamical black holes represented by our new solutions.
Our conclusions and discussions are summarized in Sec.~\ref{sec:summary}.
Einstein-Weyl gravity in arbitrary dimensions is explained in Appendix~\ref{app:Weyl}.
The details of computation to derive the Wald-Kodama dynamical entropy are presented in Appendix~\ref{app:entropy}.
Our basic notation follows~\cite{wald}.
The convention for the Riemann curvature tensor is $[\nabla _\rho ,\nabla_\sigma]V^\mu ={R^\mu }_{\nu\rho\sigma}V^\nu$ and $R_{\mu \nu }={R^\rho }_{\mu \rho \nu }$.
The Minkowski metric is taken as diag$(-,+,\cdots,+)$, and Greek indices run over all spacetime indices.
$G_n$ is the $n$-dimensional Newtonian constant.
We adopt the units such that $c=1$ and the four-dimensional Newtonian constant is described as $G(:=G_4)$.

\section{Preliminaries}
\label{sec2}
\subsection{The most general quadratic curvature gravity}
In this section, we consider the most general quadratic curvature gravity in $n(\ge 3)$ dimensions:
\begin{align}
I=& \frac{1}{16\pi G_n}\int \D^nx\,\sqrt{-g} \nonumber \\
&\times \Big(R-2\Lambda+\alpha R^2+\beta R_{\mu\nu}R^{\mu\nu}+\gamma {L}_{\rm GB}\Big)+I_{\rm m}, \label{action}
\end{align}
where $I_{\rm m}$ is the action for matter fields.
Here $\alpha$, $\beta$, and~$\gamma$ are coupling constants to the quadratic terms and ${L}_{\rm GB}$ is the Gauss-Bonnet term defined by
\begin{align}
{L}_{\rm GB} := R^2-4R_{\mu\nu}R^{\mu\nu}+R_{\mu\nu\rho\sigma}R^{\mu\nu\rho\sigma}.
\end{align}

The resulting field equations are
\begin{align}
{\cal G}_{\mu\nu}=&\ 8\pi G_n\, T_{\mu\nu}, \label{EFEs}
\end{align}
where the curvature tensor ${\cal G}_{\mu\nu}$ is defined by
\begin{align}
{\cal G}_{\mu\nu}:= G_{\mu\nu}+\Lambda g_{\mu\nu}+H_{\mu\nu}
\end{align}
and the energy-momentum tensor $T_{\mu\nu}$ for matter comes from the matter action $I_{\rm m}$.
Here $H_{\mu\nu}$ is the quadratic curvature tensor defined by
\begin{align}
H_{\mu\nu}:= \alpha H^{(1)}_{\mu\nu}+\beta H^{(2)}_{\mu\nu}+\gamma H^{(3)}_{\mu\nu},
\end{align}
where
\begin{widetext}
\begin{align}
H^{(1)}_{\mu\nu}:=&\ 2 R\biggl(R_{\mu\nu}-\frac14 g_{\mu\nu}R\biggl) +2\left(g_{\mu\nu}\dalm R-\nabla_\mu \nabla_\nu R\right), \\
H^{(2)}_{\mu\nu}:=&\ 2\biggl(R_{\mu\rho}R_{\nu}^{~~\rho}-\frac14 g_{\mu\nu}R_{\rho\sigma}R^{\rho\sigma}\biggl)+\left(\dalm R_{\mu\nu}+\frac12g_{\mu\nu}\dalm R-2\nabla_\rho\nabla_{(\mu}R_{\nu)}^{~~\rho}\right), \\
H^{(3)}_{\mu\nu}:=&\ 2\left(RR_{\mu\nu}-2R_{\mu\alpha}
R^\alpha_{~\nu}-2R^{\alpha\beta}R_{\mu\alpha\nu\beta} +R_{\mu}^{~\alpha\beta\gamma}R_{\nu\alpha\beta\gamma}\right)
-\frac12g_{\mu\nu}{L}_{\rm GB}.
\end{align}
\end{widetext}
The Gauss-Bonnet term is dynamical only for $n\ge 5$ and hence $H^{(3)}_{\mu\nu}\equiv 0$ holds for $n\le 4$.

\subsection{Spacetime metric and maximally symmetric vacua}
In the present paper, we consider the following Vaidya-type metric:
\begin{align}
\D s^2=&-f(v,r)\D v^2+2\D v\D r+r^2\gamma_{ij}\D z^i\D z^j, \label{ansatz}
\end{align}
where $i,j=2,3,\cdots,n-1$ and $\gamma_{ij}\D z^i\D z^j$ is the line element on the $(n-2)$-dimensional maximally symmetric base manifold $(K^{n-2},\gamma_{ij})$ with its curvature $k=1,0,-1$, namely the Riemann tensor ${}^{(n-2)}R^{ij}_{~~kl}$ on $(K^{n-2},\gamma_{ij})$ is given by
\begin{align}
{}^{(n-2)}R^{ij}_{~~kl}=k\,(\delta^i_k\delta^j_l-\delta^i_l\delta^j_k).
\end{align}
We assume that the base manifold $K^{n-2}$ is compact for simplicity and hereafter $V_{n-2}^{(k)}$ denotes its volume.

The metric function $f(v,r)$ for the maximally symmetric vacuum solution is given by
\begin{align}
f(r)=k-\frac{2\lambda}{(n-1)(n-2)} r^2,\label{sol-max-n}
\end{align}
where $\lambda$ is an effective cosmological constant and $k\lambda\ne 0$ is required.
The spacetime is locally Minkowski, de~Sitter (dS), and anti-de~Sitter (AdS) for $\lambda=0$, $\lambda>0$, and $\lambda<0$, respectively.
The trace of the field equations (\ref{EFEs}) gives the following algebraic equation to determine $\lambda$:
\begin{align}
&2(n-4)\lambda^2\big[(n-1)(n\alpha+\beta)+(n-2)(n-3)\gamma\big]\nonumber \\
&+(n-1)(n-2)^2(\lambda-\Lambda)=0.\label{master-dS}
\end{align}

In four dimensions ($n=4$), the quadratic terms do not affect the value of the cosmological constant and we have $\lambda=\Lambda$.
In higher dimensions, the number of real solutions of Eq.~(\ref{master-dS}) depends on the parameters and there can be two distinct maximally symmetric vacua.
However, there are two cases which admit the unique vacuum.
The first case is
\begin{align}
\gamma=&-\frac{(n-1)(n\alpha+\beta)}{(n-2)(n-3)}, \label{degenerate-cond2}
\end{align}
which gives the effective cosmological constant ${\lambda=\Lambda}$.
The other case is
\begin{align}
\Lambda=&-\frac{(n-1)(n-2)^2}{8(n-4)[(n-1)(n\alpha+\beta)+(n-2)(n-3)\gamma]}, \label{degenerate-cond1}
\end{align}
which gives $\lambda=2\Lambda$.

\subsection{Critical gravity}
Critical gravity is described by the action (\ref{action}) in four dimensions ($n=4$) with a special choice of the coupling constants~\cite{criticalgravity}.
Eliminating massive scalar mode around the AdS background requires $\beta=-3\alpha$ and the condition for stable (non-tachyonic) massive spin-2 mode is $0<\alpha\le -1/2\Lambda$ with equality giving the massless spin-2 mode.
Critical gravity is realized if spin-2 mode becomes massless, namely $\alpha= -1/2\Lambda$.
Finally, the relations between the coupling constants for critical gravity are
\begin{align}
\beta=-3\alpha=\frac{3}{2\Lambda}
\label{critgrav}
\end{align}
with $\Lambda<0$~\cite{criticalgravity}.

In higher dimensions ($n\ge 5$), the Gauss-Bonnet term comes into play and then unitary theory is realized for
\begin{align}
\beta = -\frac{4(n-1)}{n}\alpha,~~\gamma = -\frac{(n-1)(n-2)[4\alpha(n-2)\lambda+n]}{4n(n-3)(n-4)\lambda},
\end{align}
where $\lambda$ is the (negative) effective cosmological constant determined by Eq.~(\ref{master-dS})~\cite{criticalgravity-higher}.
This is the higher-dimensional counterpart of critical gravity.

In this higher-dimensional case, there are two free parameters $\alpha$ and $\Lambda$ as coupling constants.
If one assumes the condition (\ref{degenerate-cond2}) for the uniqueness of the maximally symmetric vacuum in addition, the relations between the coupling constants are
\begin{align}
\beta = -\frac{4(n-1)}{n}\alpha,~~\Lambda = -\frac{n}{8\alpha},~~\gamma = -\frac{(n-1)(n-2)}{n(n-3)}\alpha, \label{higher-critical}
\end{align}
where $\alpha$ is a single parameter as a coupling constant~\cite{criticalgravity-higher}.
Actually, critical gravity and its higher-dimensional counterpart with the relations (\ref{higher-critical}) are special classes of the Einstein-Weyl gravity with a cosmological constant~\cite{criticalgravity}. (See Appendix~\ref{app:Weyl}.)

\subsection{Definition of a dynamical black hole}
In the following sections, we will study properties of the spacetime (\ref{ansatz}) representing a dynamical black hole.
In a dynamical situation, a black hole is defined by a class of the trapping horizon~\cite{hayward1994}.
Here we summarize the definition of a dynamical black hole described by the metric (\ref{ansatz}).

In the spacetime (\ref{ansatz}), radial null geodesics satisfy
\begin{equation}
0=\D v(-f\D v+2\D r).
\end{equation}
While ingoing radial null geodesics are represented by $v=$constant, outgoing null geodesics satisfy
\begin{equation}
\frac{\D r}{\D v}=\frac12f. \label{out-null}
\end{equation}
The tangent vectors along the future-directed radial outgoing and ingoing null geodesics, which are denoted respectively as $k^\mu$ and $\l^\mu$, are given by
\begin{equation}
k^\mu\frac{\partial}{\partial x^\mu}=\frac{\partial}{\partial v}+\frac{f}{2}\frac{\partial}{\partial r},\qquad l^\mu\frac{\partial}{\partial x^\mu}=-\frac{\partial}{\partial r},\label{null-vectors}
\end{equation}
which satisfy ${k^\mu k_\mu=l^\mu l_\mu=0}$ and ${k^\mu l_\mu=-1}$.

The surface area with constant $v$ and $r$ is given by ${A:=V_{n-2}^{(k)}\,r^{n-2}}$.
The expansion along outgoing and ingoing radial null geodesics are computed as
\begin{align}
\Theta_+:=&\ \frac{k^\mu\nabla_\mu{A}}{{A}}=\frac{1}{{A}}\biggl(\frac{\partial {A}}{\partial
v}+\frac{f}{2}\frac{\partial {A}}{\partial r}\biggl)
=\frac{n-2}{2r}f,\label{out-expansion}\\
\Theta_-:=&\ \frac{l^\mu\nabla_\mu{A}}{{A}}=-\frac{1}{{A}}\frac{\partial {A}}{\partial r}
=-\frac{n-2}{r},\label{expansion2}
\end{align}
respectively, where we used Eq.~(\ref{null-vectors}).
A trapping horizon is defined by the vanishing null expansion.
In the present case, a trapping horizon is defined by ${\Theta_+=0}$, and hence its location ${r=r_{\rm h}(v)}$ is given by solving the following algebraic equation:
\begin{align}
f(v,r_{\rm h})=0.\label{t-horizon}
\end{align}

A trapping horizon is classified in the following manner; {\it future} if ${\Theta_-<0}$, {\it past} if ${\Theta_->0}$, {\it bifurcating} if ${\Theta_-=0}$, {\it outer} if ${{\cal L}_-\Theta_+<0}$, {\it inner} if ${{\cal L}_-\Theta_+>0}$, and {\it degenerate} if ${{\cal L}_-\Theta_+=0}$, where ${\cal L}_-$ is the derivative along the ingoing radial null geodesic~\cite{hayward1994}.
In this classification, a dynamical black hole is defined by a future outer trapping horizon~\cite{hayward1994}.

Equation~(\ref{expansion2}) shows that $\Theta_-<0$ is always satisfied.
On the other hand, from the following expression
\begin{align}
{\cal L}_-\Theta_+:=&l^\mu \nabla_\mu \Theta_+=-\frac{\partial \Theta_+}{\partial r}=\frac{n-2}{2r^2}\biggl(f-r\frac{\partial f}{\partial r}\biggl),
\end{align}
we obtain
\begin{align}
{\cal L}_-\Theta_+|_{r=r_{\rm h}}=-\frac{n-2}{2r}\frac{\partial f}{\partial r}\biggl|_{r=r_{\rm h}}.\label{der-expansion}
\end{align}
Therefore, a future outer trapping horizon is realized if the following condition is satisfied:
\begin{align}
\frac{\partial f}{\partial r}\biggl|_{r=r_{\rm h}}>0.
\label{xyz}
\end{align}

Because the line element along the orbit of a trapping horizon $r=r_{\rm h}(v)$ is given by
\begin{equation}
\D s^2=2\biggl(\frac{\D r_{\rm h}}{\D v}\biggl)\D v^2+r^2\gamma_{ij}\D z^i\D z^j,
\end{equation}
where we used Eq.~(\ref{out-null}), the signature of the trapping horizon is determined by the sign of $\D r_{\rm h}/\D v$, independent of the theory.
Therefore, the area of the trapping horizon increases (decreases) along its generator if and only if it is spacelike (timelike).

If a future outer trapping horizon is timelike, it does not capture a concept of a black hole as a one-way membrane and therefore it should be non-timelike in order to define a black hole in a dynamical situation appropriately.
In general relativity, for the spacetime with spherical ($k=1$), planar ($k=0$), or hyperbolic ($k=-1$) symmetry, it was shown that, under the null energy condition, a future outer (inner) trapping horizon is non-timelike (non-spacelike) and its area is non-decreasing (non-decreasing) along its generator~\cite{hayward1994,nm2008}.

\section{Charged Vaidya-type solutions}
\label{sec3}
\subsection{Matter fields}
In the present paper, we consider a null dust and a Maxwell field as matter fields, of which energy-momentum tenor is given by
\begin{align}
T_{\mu\nu}= T^{(1)}_{\mu\nu}+T^{(2)}_{\mu\nu}.
\end{align}
Here $T^{(1)}_{\mu\nu}$ is the energy-momentum tensor for a Maxwell field, given by
\begin{align}
T^{(1)}_{\mu\nu}= \frac{1}{4\pi\sigma^2}\biggl(F_{\mu\rho}F_{\nu}^{~\rho}-\frac14 g_{\mu\nu}F_{\rho\sigma}F^{\rho\sigma}\biggl),
\end{align}
where $F_{\mu\nu}:=\partial_\mu A_\nu-\partial_\nu A_\mu$ is the Faraday tensor and $\sigma$ is a coupling constant.
On the other hand, $T^{(2)}_{\mu\nu}$ is the energy-momentum tensor of a null dust fluid, given by
\begin{align}
T^{(2)}_{\mu\nu}= \rho\, l_\mu l_\nu,
\end{align}
where $\rho$ is the energy density for a null dust and $l^\mu$ is a null vector.
In the coordinate system (\ref{ansatz}), we consider the null vector $l^\mu$ for a null dust fluid in the following ingoing form:
\begin{align}
l^\mu\frac{\partial}{\partial x^\mu}=&-\frac{\partial}{\partial r},
\end{align}
which gives ${T^{(2)}}^{r}_{~v}=\rho$.

In the case of $n\ge 4$, the form of the gauge field for the electric solution is given by
\begin{align}
A_\mu \D x^\mu=&-\frac{Q_e(v)}{r^{n-3}}\D v,\label{gauge-sol}
\end{align}
where $Q_e(v)$ is a function of $v$.
This expression gives
\begin{align}
F_{vr}=&-(n-3)\frac{Q_e}{r^{n-2}},\qquad F^{vr}=(n-3)\frac{Q_e}{r^{n-2}}
\end{align}
and the only non-zero components of $\nabla_\nu F^{\mu\nu}$ is
\begin{align}
\nabla_\nu F^{r\nu}=&-(n-3)\frac{{\dot Q}_{e}}{r^{n-2}},
\end{align}
where a dot denotes the derivative with respect to $v$.

Then, the non-zero components of the energy-momentum tensor for $n\ge 4$ are
\begin{align}
T^{v}_{~v}=&\ T^{r}_{~r}=-\frac{(n-3)^2\,Q_e^2}{8\pi  \sigma^2 \,r^{2(n-2)}},\nonumber \\
T^{i}_{~j}=&\ \frac{(n-3)^2\,Q_e^2}{8\pi  \sigma^2 \,r^{2(n-2)}}\delta^i_{~j}, \qquad T^{r}_{~v}=\rho.
\end{align}
The conservation equations for the total energy-momentum tensor $\nabla_\nu T^{\nu}_{~\mu}=0$ has only one non-trivial component for $\mu=v$, which is written as
\begin{align}
\frac{(n-3)^2Q_e{\dot Q}_{e}}{4\pi  \sigma^2 \,r^{2(n-2)}}=\rho'+\frac{n-2}{r}\rho, \label{conservation-eq}
\end{align}
where a prime denotes the derivative with respect to $r$.
Now we are ready to present exact solutions in this system.

\subsection{General relativity}
The topological and arbitrary $n(\ge 4)$-dimensional generalization of the charged Vaidya solution in general relativity ($\alpha=\beta=\gamma=0$)  is given by
\begin{align}
\D s^2=&-f(v,r)\D v^2+2\D v\D r+r^2\gamma_{ij}\D z^i\D z^j, \label{metric-n}\\
A_\mu \D x^\mu=&-\frac{Q_e(v)}{r^{n-3}}\D v,\qquad l^\mu\frac{\partial}{\partial x^\mu}=-\frac{\partial}{\partial r} \label{matter-n}
\end{align}
with
\begin{align}
f(v,r)=&\ k-\frac{2m(v)}{r^{n-3}}+\frac{2(n-3)G_n Q_e(v)^2}{(n-2) \sigma^2 \,r^{2(n-3)}}-{\bar \Lambda}r^2,\label{sol-charged-Vaidya}\\
\rho(v,r)=&\ \frac{n-2}{8\pi G_n \,r^{n-2}}\biggl({\dot m}-\frac{2(n-3) G_n Q_e{\dot Q}_{e}}{(n-2) \sigma^2 \,r^{n-3}}\biggl), \label{sol-charged-Vaidya-rho}
\end{align}
where ${\bar \Lambda}:=2\Lambda/(n-1)(n-2)$~\cite{chargedVaidya-GR}. (See also~\cite{po2006,opz2008,f1974}.)
Here $m(v)$ and $Q_e(v)$ are arbitrary functions.
The expression (\ref{sol-charged-Vaidya-rho}) is consistent with Eq.~(\ref{conservation-eq}), of course.

\subsection{Einstein-Gauss-Bonnet gravity}
In Einstein-Gauss-Bonnet gravity (${\alpha=\beta=0}$) for ${n\ge 5}$, there is the charged Vaidya-type solution given by Eqs.~(\ref{metric-n}) and (\ref{matter-n}) with Eq.~(\ref{sol-charged-Vaidya-rho}) and the following metric function:
\begin{widetext}
\begin{align}
f(v,r)=&k+\frac{r^2}{2{\bar\gamma}}\biggl(1\mp\sqrt{1+4{\bar \gamma}{\bar\Lambda}+8{\bar\gamma}\frac{m(v)}{r^{n-1}}-8{\bar\gamma}\frac{(n-3) G_n Q_e(v)^2}{(n-2) \sigma^2 \,r^{2(n-2)}}}\biggl),\label{sol-GB-Vaidya}
\end{align}
\end{widetext}
where ${\bar\gamma}:=(n-3)(n-4)\gamma$.
This solution with ${k=1}$ has been obtained in~\cite{chargedVaidya-EGB} in a more general context.

There are two branches of solutions corresponding to the sign in the metric function (\ref{sol-GB-Vaidya}).
While the solution with a minus sign (GR branch) admits the general-relativistic limit $\gamma\to 0$, the solution with a plus sign (non-GR branch) is diverging in this limit.
For $\gamma>0$, the maximally symmetric vacuum solution does not suffer from the ghost instability only in the GR branch~\cite{Zwiebach:1985uq,bb1996}.

\subsection{New solutions in four-dimensional quadratic curvature gravity}
\label{sec:newsolution}
In four dimensions ($n=4$),  the Gauss-Bonnet term becomes topological and therefore the most general quadratic curvature gravity becomes much simpler.
In this case, new charged Vaidya-type exact solutions are available.

\subsubsection{ Case $\beta = 0$}
First let us consider the case with $\beta=0$.
The theory is then a special class of the so-called $F(R)$ gravity~\cite{f(R)}, of which action and the field equations are
\begin{align}
&I=\frac{1}{16\pi G}\int \D^4x\,\sqrt{-g}\,F(R)+I_{\rm m},\label{F(R)action}\\
&\frac{\D F}{\D R}R_{\mu\nu}-\frac12Fg_{\mu\nu}-(\nabla_\mu\nabla_\nu-g_{\mu\nu}\dalm)\frac{\D F}{\D R}=8\pi G\, T_{\mu\nu}. \label{F(R)-eq}
\end{align}
Our quadratic curvature theory (\ref{action}) with ${\beta=0}$ corresponds to ${F(R)=R-2\Lambda+\alpha R^2}$.
For ${\beta=0}$ with ${1+8\alpha\Lambda \ne 0}$, there is an exact solution given by Eqs.~(\ref{metric-n}) and (\ref{matter-n}) with
\begin{align}
f(v,r)=&\ k-\frac{2m(v)}{r}+\frac{G Q_e(v)^2}{(1+8\alpha\Lambda) \sigma^2 r^2}-\frac{\Lambda}{3}r^2,\label{sol-4dim-3} \\
\rho(v,r)=&\ \frac{1}{4\pi G \,r^2}\bigg[(1+8\alpha\Lambda){\dot m}-\frac{G Q_e{\dot Q}_{e}}{ \sigma^2 \,r}\bigg],\label{sol-4dim-3-rho}
\end{align}
where $m(v)$ and $Q_e(v)$ are arbitrary functions.

For ${\beta=0}$ with ${1+8\alpha\Lambda =0}$, on the other hand, there is the following dynamical solution in vacuum (${\rho=Q_e=0}$):
\begin{align}
f(v,r)= k-\frac{2m(v)}{r}+\frac{w(v)}{r^2}-\frac{\Lambda}{3}r^2,\label{sol-4dim-4}
\end{align}
where $m(v)$ and $w(v)$ are arbitrary functions.
With ${\beta=0}$ and $1+8\alpha\Lambda =0$, the function $F(R)$ in the action (\ref{F(R)action}) becomes a perfect square:
\begin{align}
F(R)=-\frac{(R-4\Lambda)^2}{8\Lambda}.
\end{align}
Thus, not only the spacetime (\ref{ansatz}) with the metric function (\ref{sol-4dim-4}), but also {\it any} metric satisfying a single scalar equation $R=4\Lambda$ is an exact vacuum solution of the field equations (\ref{F(R)-eq}) in this theory.
For this reason, the theory with $\beta=0$ and $1+8\alpha\Lambda =0$ is singular among all the quadratic curvature theories in four dimensions.

\subsubsection{ Case $\beta \ne 0$}
Lastly let us consider the case of $\beta\ne 0$.
In this case, an exact solution is given by Eqs.~(\ref{metric-n}) and (\ref{matter-n}) with
\begin{align}
f(v,r)=&\ k-\frac{2m(v)}{r}-\frac{\Lambda}{3}r^2,\label{sol4dim-1}\\
\rho(v,r)=&\ \frac{1}{4\pi G \,r^2}\biggl[\big[1+2\Lambda(4\alpha+\beta)\big]{\dot m}-\frac{2\beta}{r}{\ddot m}\biggl],\label{sol4dim-1-rho}
\end{align}
where $m(v)$ and $Q_e(v)$ satisfy
\begin{align}
{\dot m}=\frac{G Q_e^2}{4  \sigma^2 \beta}.\label{mdot}
\end{align}
In the limit of $\beta\to 0$, this solution reduces to the solution (\ref{sol-4dim-3}) with $Q_e=0$.
In this sense, the solutions (\ref{sol-4dim-3}) and (\ref{sol4dim-1}) are two different branches of the charged solutions.

In particular, critical gravity is realized for the coupling constants satisfying (\ref{critgrav}), implying ${1+2\Lambda(4\alpha+\beta)=0}$, ${\beta<0}$.

\section{Black-hole dynamics in the new solutions}
\label{sec4}
In the spacetime (\ref{ansatz}), a future outer trapping horizon ${r=r_{\rm h}}$ is given by $f(v,r_{\rm h})=0$ with $\partial f/\partial r|_{r=r_{\rm h}}>0$.
In general relativity, under the null energy condition, a future outer trapping horizon is non-timelike and its area is non-decreasing along its generator~\cite{hayward1994,nm2008}.
Therefore, if the energy density of the null dust fluid (\ref{sol-charged-Vaidya-rho}) is non-negative, the null energy condition is satisfied for the total energy momentum tensor $T_{\mu\nu}$ and then no pathological behavior is observed for dynamical black holes.
In the charged Vaidya-type solution (\ref{metric-n}) and its generalization in Einstein-Gauss-Bonnet gravity (\ref{sol-GB-Vaidya}), the energy density of a null dust is non-negative in the whole spacetime if ${\dot m}\ge 0$ and $\partial_v(Q_{\rm e}^2)\le 0$ hold.

However, different from general relativity, a pathological behavior of dynamical black holes is observed in in Einstein-Gauss-Bonnet gravity.
In Einstein-Gauss-Bonnet gravity, a future outer trapping horizon is non-timelike (non-spacelike) and its area is non-decreasing (non-increasing) along its generator in the GR branch (non-GR branch) under the null energy condition~\cite{nm2008,maeda2010}.
This clearly shows that the non-GR branch is pathological and only the solutions in the GR branch are well-behaving.

Nevertheless, it was shown that the Wald-Kodama dynamical entropy of the future outer trapping horizon, given by
\begin{align}
S_{\rm WK}=&\frac{A_{\rm h}}{4G_n}\bigg[1+2(n-2)(n-3)\frac{k\gamma}{r_{\rm h}^2}\bigg],
\end{align}
where ${A_{\rm h}:=V^{(k)}_{n-2}\,r_{\rm h}^{n-2}}$ is the area of the trapping horizon, is non-decreasing along its generator in both branches under the null energy condition~\cite{nm2008,maeda2010}.
(See Appendix~\ref{app:entropy} for the definition of the Wald-Kodama dynamical entropy.)
These results imply that the entropy-increasing law is more fundamental than the area-increasing law in black-hole physics.

The universality of the entropy-increasing law rather than the area-increasing law has been observed also in the Vaidya-type solution in BHT massive gravity in three dimensions with a unique AdS vacuum~\cite{maeda2011}.
However, it has been reported that the Wald-Kodama entropy of a dynamical black hole can decrease in BHT massive gravity in a vacuum solution with a certain relation between the coupling constants~\cite{fs2013,flory-thesis}.
Now let us see the properties of a dynamical black hole represented by new solutions obtained in section~\ref{sec:newsolution} in the most general quadratic curvature gravity in four dimensions.

\begin{table}[htb]
\begin{center}
\caption{\label{table:beta=0} Properties of a future outer trapping horizon for $\beta=0$ with $\rho\ge 0$ in the solution (\ref{sol-4dim-3}) for $1+8\alpha\Lambda\ne 0$ and the solution (\ref{sol-4dim-4}) for $1+8\alpha\Lambda=0$.}
\begin{tabular}{|c||c|c|c|c|}
\hline \hline
 $1+8\alpha\Lambda$ & ${\dot A}_{\rm h}$ & $S_{\rm WK}$ & ${\dot S}_{\rm WK}$    \\\hline\hline
$+$ & Non-negative & Positive & Non-negative \\ \hline
$0$ & Depends on $m(v)$ and $w(v)$ & $0$ & $0$ \\ \hline
$-$ & Non-positive & Negative & Non-negative \\
\hline \hline
\end{tabular}
\end{center}
\end{table}

\subsection{ Case $\beta=0$}

First let us study the solution for ${\beta=0}$.
In the case with ${1+8\alpha\Lambda \ne 0}$, the location of the trapping horizon ${r=r_{\rm h}(v)}$ is determined by ${f(v,r_{\rm h})=0}$ with the metric function (\ref{sol-4dim-3}), namely
\begin{align}
0=k-\frac{2m(v)}{r_{\rm h}}+\frac{G Q_e(v)^2}{(1+8\alpha\Lambda) \sigma^2 r_{\rm h}^2}-\frac{\Lambda}{3}r_{\rm h}^2.
\end{align}
The number of the future outer trapping horizons depends on the values of $m$ and $Q_e$ and is shown by the result in~\cite{tm2005}.
Differentiating the above equation with respect to $v$, we obtain
\begin{align}
0= -\frac{2{\dot m}}{r_{\rm h}}+\frac{2G\,Q_e\dot Q_e}{(1+8\alpha\Lambda) \sigma^2 r_{\rm h}^2}+\frac{\D r_{\rm h}}{\D v}\frac{\partial f}{\partial r}\biggl|_{r=r_{\rm h}},
\end{align}
which gives
\begin{align}
\frac{\D r_{\rm h}}{\D v}\frac{\partial f}{\partial r}\biggl|_{r=r_{\rm h}}=
\frac{8\pi G\, r_{\rm h}}{1+8\alpha\Lambda}\,\rho(v,r_{\rm h}).
\end{align}
Because $\partial f/\partial r|_{r=r_{\rm h}}>0$ holds on a future outer trapping horizon, see~(\ref{xyz}), the above equation shows that its \emph{area} is non-decreasing (non-increasing) for $1+8\alpha\Lambda>(<)0$ under the condition $\rho \ge 0$.
This shows a pathological behavior of a dynamical black hole for $1+8\alpha\Lambda<0$.

\begin{table*}[htb]
\begin{center}
\caption{\label{table:betanot0} Properties of a future outer trapping horizon for $\beta\ne 0$ with $\rho\ge 0$ in the solution (\ref{sol4dim-1}). Simple criteria are not obtained in the blank slots.}
\begin{tabular}{|c|c||c|c|c|}\hline\hline
$~~\beta~~$ & $1+2\Lambda(4\alpha+\beta)$ & ${\dot A}_{\rm h}$  & $S_{\rm WK}$  & ${\dot S}_{\rm WK}$  \\ \hline\hline
 & $+$ & Positive &  &  Non-negative for ${\ddot m}\le 0$ and $\Lambda\le 0$   \\ \cline{2-5}
$+$ & $0$ & Positive & Negative & Non-negative for ${\ddot m}\le 0$ and $\Lambda\le 0$ \\ \cline{2-5}
 & $-$ & Positive & Negative  &  \\ \hline
 & $+$ & Negative &  &     \\ \cline{2-5}
$-$ & $0$ & Negative & Negative  &  Non-negative for ${\ddot m}\ge 0$ and $\Lambda\ge 0$  \\ \cline{2-5}
 & $-$ & Negative &Negative  & Non-negative for ${\ddot m}\ge 0$ and $\Lambda\ge 0$  \\ \hline\hline
\end{tabular}
\end{center}
\end{table*}

Now let us see whether such a pathological behavior is observed in the \emph{dynamical entropy}.
The Wald-Kodama dynamical entropy of this black hole is given by
\begin{align}
S_{\rm WK}=\frac{1+8\alpha\Lambda}{4G}\,A_{\rm h}. \label{S-beta=0}
\end{align}
(See Appendix~\ref{app:entropy1} for derivation.)
The above expression shows that the dynamical entropy of the future outer trapping horizon is positive (negative) for ${1+8\alpha\Lambda>(<)0}$.
However, the differentiation of Eq.~(\ref{S-beta=0}) with respect to $v$ gives
\begin{align}
{\dot S}_{\rm WK}=\frac{1+8\alpha\Lambda}{4G}\,{\dot A}_{\rm h},
\end{align}
which shows that the entropy is non-decreasing under the condition ${\rho \ge 0}$ independent of the sign of ${1+8\alpha\Lambda}$.
This result provides another example that the entropy-increasing law is more fundamental than the area-increasing law.

In the singular theory with $\beta=0$ and $1+8\alpha\Lambda =0$, on the other hand, the vacuum solution (\ref{sol-4dim-4}) contains two arbitrary functions $m(v)$ and $w(v)$.
Therefore, the area of the future outer trapping horizon can decrease even in vacuum by choosing these functions appropriately.
However, in contrast to the case of $1+8\alpha\Lambda \ne 0$, the dynamical entropy (\ref{S-beta=0}) of the black hole is identically zero, independent of $m(v)$ and $w(v)$.
The results in this subsection are summarized in Table~\ref{table:beta=0}.

\subsection{ Case ${\beta \ne 0}$}
Next let us study the solution for ${\beta\ne 0}$, of which metric function is given by Eq.~(\ref{sol4dim-1}).
Since Eq.~(\ref{mdot}) shows that ${Q_e(v)\equiv 0}$ gives the topological generalization of the Schwarzschild-(A)dS vacuum solution, we consider the case of ${Q_e(v)\ne 0}$ and focus on the situation where ${\rho\ge 0}$ is satisfied everywhere.
For ${\beta>(<)0}$, ${{\dot m}>(<)0}$ holds by Eq.~(\ref{mdot}), and therefore ${\rho\ge 0}$ is satisfied everywhere if ${1+2\Lambda(4\alpha+\beta)\ge (\le)0}$ and ${\ddot m}\le(\ge) 0$ holds.
Because ${{\dot m}>(<)0}$ implies that the area of the future outer trapping horizon increases (decreases), the area-increasing law is satisfied (violated) for ${\beta>(<)0}$.

The Wald-Kodama dynamical entropy of this black hole is given by
\begin{align}
S_{\rm WK}=&\,\Big[1+2\Lambda(4\alpha+\beta)\Big]\frac{A_{\rm h}}{4G}-\frac{\beta V^{(k)}_2{\dot m}}{G\kappa_{\rm TH}\, r_{\rm h}} \nonumber \\
=&\ \Big[1+2\Lambda(4\alpha+\beta)\Big]\frac{A_{\rm h}}{4G}-\frac{2\beta V^{(k)}_2{\dot m}}{G(k-\Lambda r_{\rm h}^2)},\label{S-sol2}
\end{align}
where ${\kappa_{\rm TH}(>0)}$ is the surface gravity on the trapping horizon.
(See Appendix~\ref{app:entropy2} for derivation.)
The second term is non-positive by Eq.~(\ref{mdot}) and therefore Wald-Kodama entropy is negative in the case of ${1+2\Lambda(4\alpha+\beta)\le 0}$.

Actually, absence of a massive scalar mode around the AdS background requires ${\beta=-3\alpha}$, which gives ${1+2\Lambda(4\alpha+\beta)=1+2\Lambda \alpha}$.
Therefore, the inequality ${0<\alpha\le -1/2\Lambda}$, the condition for stable (non-tachyonic) massive spin-2 mode, ensures the non-negativity of the first term in Eq.~(\ref{S-sol2}).

Next let us see the time-evolution of the Wald-Kodama entropy.
Differentiating Eq.~(\ref{S-sol2}) with respect to $v$, we obtain
\begin{align}
{\dot S}_{\rm WK}=&\Big[1+2\Lambda(4\alpha+\beta)\Big]\frac{{\dot A}_{\rm h}}{4G}-\frac{2\beta V^{(k)}_2{\ddot m}}{G(k-\Lambda r_{\rm h}^2)} \nonumber \\
&-\frac{2\beta \Lambda {\dot m}{\dot A}_{\rm h}}{G(k-\Lambda r_{\rm h}^2)^2}. \label{dot-S}
\end{align}
By $\kappa_{\rm TH}>0$ and Eq.~(\ref{kappa-TH}), $k-\Lambda r_{\rm h}^2>0$ holds for a future outer trapping horizon.

In the case of ${\beta>0}$, Eq.~(\ref{mdot}) shows ${{\dot m}>0}$ (and hence ${{\dot A}_{\rm h}>0}$) and so we assume ${1+2\Lambda(4\alpha+\beta)\ge 0}$ and ${{\ddot m}\le }0$ in order that $\rho\ge 0$ is satisfied everywhere.
Then, because the first and the second terms in Eq.~(\ref{dot-S}) are non-negative, the entropy is non-decreasing for $\Lambda\le 0$.

On the other hand, in the case of ${\beta<0}$, Eq.~(\ref{mdot}) shows ${{\dot m}<0}$ (and hence ${{\dot A}_{\rm h}<0}$) and hence we assume ${1+2\Lambda(4\alpha+\beta)\le 0}$ and ${{\ddot m}\ge 0}$ for ${\rho\ge 0}$ everywhere.
Then, because the first and the second terms in Eq.~(\ref{dot-S}) are non-negative, the entropy is non-decreasing for ${\Lambda\ge 0}$.
The results obtained up to here are summarized in Table~\ref{table:betanot0}.

Now let us focus on \emph{critical gravity}, which is realized for ${1+2\Lambda(4\alpha+\beta)=0}$, ${\beta<0}$, and ${\Lambda<0}$.
In this case, ${{\dot m}<0}$ holds by Eq.~(\ref{mdot}), which shows that the area of the future outer trapping horizon is decreasing and the Wald-Kodama entropy (\ref{S-sol2}) is negative.
Even worse, it is shown that the Wald-Kodama entropy can decrease under the null energy condition.
By Eq.~(\ref{sol4dim-1-rho}), ${\rho\ge 0}$ requires ${{\ddot m}\ge 0}$.
Then we consider the mass function ${m(v)=m_0+m_1v}$ with ${m_0>0}$ and ${m_1<0}$.
With this mass function, there is a certain domain of $v$ admitting a future outer trapping horizon.
In this case, we have ${{\ddot m}=0}$ and ${\rho=0}$, so that the Wald-Kodama entropy is negative and decreasing only with a Maxwell field.
These results certainly show pathological aspects of the theory at the nonlinear level.

\section{Summary and discussions}
\label{sec:summary}
In the present paper, we have obtained Vaidya-type exact solutions in the most general quadratic curvature gravity in four dimensions in the presence of a null dust fluid and a Maxwell field.
These solutions represent a dynamical black hole defined by a future outer trapping horizon and we have studied their physical properties in the case with the positive energy density of the null dust.

Our main result is provided in critical gravity, which is the most important class of the theories in the present system.
Our solution shows that the area of a future outer trapping horizon is decreasing by the injection of matter fields with positive energy density.
This is never realized in general relativity and surely shows that a black hole does not capture a concept of a one-way membrane under the null energy condition in critical gravity.

Indeed, the violation of the area-increasing law has been observed also in our solutions with different coupling constants in the theory and such examples are known in Einstein-Gauss-Bonnet gravity~\cite{nm2008} and BHT massive gravity with a unique AdS vacuum~\cite{maeda2011}.
In these cases, however, the Wald-Kodama dynamical entropy of a black hole is non-decreasing under the null energy condition.
In the case of critical gravity, in contrast, our solution shows that the Wald-Kodama entropy is negative and can decrease.
These properties expose the pathological aspects of critical gravity at the non-perturbative level.

Then a natural question is whether the same is true or not in the higher-dimensional generalization of critical gravity~\cite{criticalgravity-higher}.
This is quite nontrivial and worth pursuing because the Gauss-Bonnet term plays into the game in higher dimensions.
We will address these problems elsewhere.

\acknowledgments
The authors thank Nobuyoshi Ohta for useful comments on renormalizability of the theory.
H.~M.~is grateful to Mario Flory for helpful correspondence.
H.~M.~thanks the Institute of Theoretical Physics in Charles University, Prague, for a kind hospitality while a part of this work was carried out. It was supported by the Czech Science Foundation Grant No. GA\v{C}R 17-01625S.

\appendix

\section{Einstein-Weyl gravity}
\label{app:Weyl}
The Weyl tensor in $n(\ge 3)$ dimensions is defined by
\begin{align}
C_{\mu\nu\rho\sigma}=&\ R_{\mu\nu\rho\sigma}-\frac{2}{n-2}(g_{\mu[\rho}R_{\sigma]\nu}-g_{\nu[\rho}R_{\sigma]\mu}) \nonumber \\
&+\frac{2}{(n-1)(n-2)}R\,g_{\mu[\rho}g_{\sigma]\nu}.
\end{align}
From this expression, the quadratic Weyl invariant is computed to give
\begin{align}
C_{\mu\nu\rho\sigma}C^{\mu\nu\rho\sigma}=&-\frac{n(n-3)}{(n-1)(n-2)}R^2 \nonumber \\
&+\frac{4(n-3)}{n-2}R_{\mu\nu}R^{\mu\nu}+L_{\rm GB}.
\end{align}
This shows that under the following relations between the coupling constants:
\begin{align}
\alpha=-\frac{n(n-3)}{(n-1)(n-2)}\,\gamma,\quad \beta=\frac{4(n-3)}{n-2}\,\gamma,
\end{align}
the action (\ref{action}) reduces to the Einstein-Weyl gravity with a cosmological constant:
\begin{align}
I=& \frac{1}{16\pi G_n}\int \D^nx\,\sqrt{-g}\Big(R-2\Lambda+\gamma C_{\mu\nu\rho\sigma}C^{\mu\nu\rho\sigma}\Big)+I_{\rm m}. \label{action-Weyl}
\end{align}

\section{Wald-Kodama entropy}
\label{app:entropy}
In this appendix, we derive the Wald-Kodama dynamical entropy for the dynamical black holes represented by our new solutions.
In the $n$-dimensional stationary spacetime generated by a Killing vector $\xi^\mu$, the Wald entropy $S$ is defined for a Killing horizon associated with $\xi^\mu$.
The Wald-Kodama dynamical entropy is defined by the same formula of the Wald entropy but with the Kodama vector $K^\mu$ instead of $\xi^\mu$.

The Kodama vector, originally introduced in the four-dimensional spherically symmetric spacetime~\cite{kodama1980}, may be defined in a more general spacetime which is a warped-product manifold ${\cal M}^n\approx M^2\times K^{n-2}$, where $(M^2,g_{ab})$ is an arbitrary two-dimensional Lorentzian spacetime and $(K^{n-2},\gamma_{ij})$ is an $(n-2)$-dimensional maximally symmetric space.
The most general metric on $({\cal M}^n, g_{\mu\nu})$ is written as
\begin{align}
\D s^2=g_{ab}(y)\D y^a \D y^b +r(y)^2\gamma_{ij}\D z^i\D z^j
\end{align}
and the Kodama vector $K^a$ is a vector on $(M^2,g_{ab})$ defined by
\begin{align}
K^{a}:=&-\varepsilon^{ab}D_b r,
\end{align}
where $D_a$ is a covariant derivative on $(M^2,g_{ab})$ and $\varepsilon_{ab}$ is a volume element on $(M^2,g_{ab})$, which satisfies $\varepsilon_{ab}\varepsilon^{ab}=-2$.
In the static case, the Kodama vector coincides with the Killing vector $\xi^\mu$ generating the symmetry of staticity.

The Kodama vector satisfies
\begin{align}
K_{a}K^a=-(Dr)^2,
\end{align}
where $(Dr)^2:=(D_ar)(D^ar)$, so that it is timelike in the untrapped region where $(Dr)^2<0$ holds and null on the trapping horizon given by $(Dr)^2=0$.
For this reason, the Kodama vector is a vector field generating a preferred time-evolution in the untrapped region and plays the same role as the horizon-generating Killing vector for a trapping horizon.

Let us consider a general gravitation theory, of which action $I$ is given by
\begin{align}
I=&\int\varepsilon_{\mu_1\cdots \mu_{n}}\,{\cal L}(g^{\mu\nu},R_{\mu\nu\rho\sigma})\,\D x^{\mu_1}\wedge \cdots \wedge \D x^{\mu_n},
\end{align}
where $\varepsilon_{\mu_1\cdots \mu_{n}}\D x^{\mu_1}\wedge \cdots \wedge \D x^{\mu_n}$ is the volume $n$-form.
The Wald entropy is defined by the following integral performed on the $(n-2)$-dimensional spacelike bifurcation surface $\Sigma$~\cite{wald1993,iyerwald1994,jkm1994}:
\begin{align}
S:=&\frac{2\pi}{\kappa} \oint {\bf Q},\label{WaldS}
\end{align}
where $\kappa$ is the surface gravity of the Killing horizon, defined by $\xi^\nu \nabla_\nu \xi_\mu =\kappa \,\xi_\mu$ evaluated on the Killing horizon.
${\bf Q}$ is the Noether charge $(n-2)$-form defined by
\begin{align}
&{\bf Q}:=\frac12 \varepsilon_{\mu\nu\alpha_1\cdots \alpha_{n-2}}\,Q^{\mu\nu}\D x^{\alpha_1}\wedge \cdots \wedge \D x^{\alpha_{n-2}},\\
&Q^{\mu\nu}:=-2X^{\mu\nu\rho\sigma}\,\nabla_\rho \xi_\sigma+4\xi_\sigma \nabla_\rho X^{\mu\nu\rho\sigma},\label{Q} \\
&X^{\mu\nu\rho\sigma}:=\frac{\partial {\cal L}}{\partial R_{\mu\nu\rho\sigma}}.
\end{align}

As shown in section 15.4 in~\cite{pad}, the quantity $\nabla_\rho X^{\mu\nu\rho\sigma}$ in Eq.~(\ref{Q}) is identically zero in Lovelock gravity~\cite{lovelock}, which is the most general second-order quasi-linear theory of gravity in arbitrary dimensions including general relativity and Einstein-Gauss-Bonnet gravity as special cases.
On the other hand, in the theories different from Lovelock gravity, which contain higher-derivative terms in the field equations, $\nabla_\rho X^{\mu\nu\rho\sigma}$ is non-vanishing in general.
Nevertheless, the second term in Eq.~(\ref{Q}) does not make any contribution to the Wald entropy because $\xi^\mu=0$ is satisfied on the bifurcation surface $\Sigma$.

The definition of the Wald-Kodama dynamical entropy is Eq.~(\ref{WaldS}) with the Kodama vector $K^a$ instead of $\xi^\mu$, integrated over the trapping horizon~\cite{hma1999}.
$\kappa$ in Eq.~(\ref{WaldS}) is then the dynamical surface gravity $\kappa_{\rm TH}$ of the trapping horizon, defined by $K^b D_{[b} K_{a]} =\kappa_{\rm TH} K_a$ evaluated on the trapping horizon.

Our Lagrangian density in four dimensions ($n=4$) is given by
\begin{align}
{\cal L}=&\frac{1}{16\pi G}(R-2\Lambda+\alpha R^2+\beta R_{\rho\sigma}R^{\rho\sigma}).
\end{align}
From the following expressions
\begin{align}
\frac{\partial R}{\partial R_{\mu\nu\rho\sigma}}=&\ g^{\nu[\sigma}g^{\rho]\mu},\\
\frac{\partial (R_{\alpha\beta}R^{\alpha\beta})}{\partial R_{\mu\nu\rho\sigma}}=&\ 2g^{\alpha[\nu}g^{\mu][\rho}g^{\sigma]\beta}R_{\alpha\beta},
\end{align}
we obtain
\begin{align}
X^{\mu\nu\rho\sigma}=&\frac{1}{16\pi G}\biggl[(1+2\alpha R)\frac{\partial R}{\partial R_{\mu\nu\rho\sigma}}+\beta\frac{\partial (R_{\alpha\beta}R^{\alpha\beta})}{\partial R_{\mu\nu\rho\sigma}}\biggl] \nonumber \\
=&\frac{1}{16\pi G}\biggl[(1+2\alpha R)g^{\nu[\sigma}g^{\rho]\mu}+2\beta g^{\alpha[\nu}g^{\mu][\rho}g^{\sigma]\beta}R_{\alpha\beta}\biggl].
\end{align}

Our four-dimensional spacetime (${n=4}$) is given by Eq.~(\ref{ansatz}), namely
\begin{align}
\D s^2=-f(v,r)\D v^2+2\D v\D r+r^2\gamma_{ij}\D z^i\D z^j.
\end{align}
In this coordinate system, we have ${\varepsilon_{vr}=1}$ and ${\varepsilon^{vr}=-1}$ and therefore the only non-zero component of the Kodama vector is ${K^v=-\varepsilon^{vr}=1}$.
The surface gravity of the trapping horizon is then expressed as
\begin{align}
\kappa_{\rm TH}=\frac12\frac{\partial f}{\partial r}\biggl|_{r=r_{\rm h}},
\end{align}
so that $\kappa_{\rm TH}>0$ for the future outer trapping horizon.

The only non-zero component of $Q^{\mu\nu}$ is
\begin{widetext}
\begin{align}
Q^{vr}=&-Q^{rv}=-2f'X^{vrvr}+4(\nabla_\rho{X^{vr\rho r}}+f\nabla_\rho{X^{vrv\rho}}),
\end{align}
where
\begin{align}
X^{vrvr}=&\ \frac{1}{16\pi G}\biggl\{-\frac{1}{2}+\alpha\Big[f''+\frac{4}{r}f'+\frac{2}{r^2}(f-k)\Big]  +\beta\Big[\frac{1}{2}f''+\frac{1}{r}f'\Big]\biggl\},\nonumber\\
\nabla_\rho{X^{vr\rho r}}=&\ \partial_v{X^{vrvr}}+\frac{\beta\,\dot f}{16\pi G\,r^2},\qquad
\nabla_\rho{X^{vrv\rho}}= \partial_r{X^{vrvr}}+\frac{\beta}{16\pi G \,r^3}
\Big[\frac{1}{2}f''r^2+(k-f)\Big],
\end{align}
and thus
\begin{align}
Q^{vr}=&\ \frac{1}{16\pi G}\biggl\{f'-\frac{\alpha}{r^3}\Big[2r^3(f'f''-2ff'''-2{\dot f}'')+8r^2(-2ff''+{f'}^2-2{\dot f}')+4r(3ff'-kf'-2{\dot f})+16f(f-k)\Big] \nonumber \\
&\hspace{16mm} -\frac{\beta}{r^3}\Big[r^3(f'f''-2ff'''-2{\dot f}'')+2r^2(-3ff''+{f'}^2-2{\dot f}')+4r(ff'-{\dot f})+4f(f-k)\Big]\biggl\}.
\end{align}

Evaluating this quantity on the trapping horizon (defined by $f(v,r_{\rm h})=0$) under the assumption that derivatives of $f$ are finite on the horizon, we obtain
\begin{align}
Q^{vr}|_{r=r_{\rm h}}=&\ \frac{1}{16\pi G}\biggl\{f'-\frac{1}{r^3}\Big[(2\alpha+\beta) r^3(f'f''-2{\dot f}'')
+2(4\alpha+\beta)r^2({f'}^2-2{\dot f}')
-4(2\alpha+\beta) r{\dot f}-4\alpha kr f'\Big]\biggl\}\biggl|_{r=r_{\rm h}}.
\end{align}
In the static case, this reduces to
\begin{align}
Q^{vr}|_{r=r_{\rm h}}=&\ \frac{f'}{16\pi G}\biggl\{1-\frac{1}{r^3}\Big[(2\alpha+\beta) r^3f''  +2(4\alpha+\beta)r^2{f'}-4\alpha kr \Big]\biggl\}\biggl|_{r=r_{\rm h}}.
\end{align}
\end{widetext}
These expressions are not much illuminating, and so direct calculations are required to derive, for each solution separately, the expression of the Wald-Kodama dynamical entropy
\begin{align}
S_{\rm WK}=&\ \frac{2\pi}{\kappa_{\rm TH}} \oint \varepsilon_{vrij}\,Q^{vr}|_{r=r_{\rm h}} \D x^{i}\wedge \D x^{j} \nonumber \\
=&\ \frac{2\pi}{\kappa_{\rm TH}} Q^{vr}|_{r=r_{\rm h}} \oint   \varepsilon_{vrij}\,2!\D z^i\D z^j \nonumber \\
=&\ \frac{2\pi}{\kappa_{\rm TH}} Q^{vr}|_{r=r_{\rm h}}\oint  \sqrt{-\det g}\, 2!\D z^i\D z^j \nonumber \\
=&\ \frac{2\pi}{\kappa_{\rm TH}} Q^{vr}|_{r=r_{\rm h}}\oint r_{\rm h}^2\sqrt{\det \gamma}\, 2!\D z^i\D z^j \nonumber \\
=&\ \frac{2\pi}{\kappa_{\rm TH}} r_{\rm h}^2\,Q^{vr}|_{r=r_{\rm h}}\oint \varepsilon_{ij}\,\D z^i\wedge \D z^j \nonumber \\
=&\ \frac{2\pi}{\kappa_{\rm TH}} r_{\rm h}^2\,Q^{vr}|_{r=r_{\rm h}}\times V^{(k)}_2,
\label{Swkeval}
\end{align}
where $V^{(k)}_2$ is the volume of the two-dimensional base manifold $(K^2,\gamma_{ij})$, which is $4\pi$ for $k=1$.

\subsection{ Case $\beta=0$}
\label{app:entropy1}
The metric function of the new solution for $\beta=0$ is given by Eq.~(\ref{sol-4dim-3}), namely
\begin{align}
f(v,r)=k-\frac{2m(v)}{r}+\frac{q(v)^2}{(1+8\alpha\Lambda)\,r^2}-\frac{\Lambda}{3}r^2,
\end{align}
where
\begin{align}
q(v)^2:=\frac{G }{ \sigma^2 }\,Q_e(v)^2.
\end{align}
The areal radius of the trapping horizon $r=r_{\rm h}$ is determined by $f(v,r_{\rm h})=0$, namely
\begin{align}
k-\frac{2m(v)}{r_{\rm h}}+\frac{q(v)^2}{(1+8\alpha\Lambda)r_{\rm h}^2}-\frac{\Lambda}{3}r_{\rm h}^2=0. \label{TH-relation2}
\end{align}
The dynamical surface gravity of the trapping horizon is given by
\begin{align}
\kappa_{\rm TH}=&\ \frac{m(v)}{r_{\rm h}^2}-\frac{q(v)^2}{(1+8\alpha\Lambda)r_{\rm h}^3}-\frac{\Lambda}{3}r_{\rm h} \nonumber \\
=&\ \frac{1}{2r_{\rm h}}\biggl[k-\Lambda r_{\rm h}^2-\frac{q(v)^2}{(1+8\alpha\Lambda)r_{\rm h}^2}\biggl],
\end{align}
where we used Eq.~(\ref{TH-relation2}) at the last equality.
With the metric function (\ref{sol-4dim-3}), $Q^{vr}$ reduces to
\begin{align}
Q^{vr}=\frac{(1+8\alpha\Lambda)(3m-\Lambda r^3)}{24\pi G\,r^2}-\frac{q^2}{8\pi G \,r^3},
\end{align}
of which value on the trapping horizon is
\begin{align}
Q^{vr}|_{r=r_{\rm h}}=&\ \frac{1+8\alpha\Lambda}{16\pi G\, r_{\rm h}}\biggl[k-\Lambda r_{\rm h}^2-\frac{q^2(v)}{(1+8\alpha\Lambda)r_{\rm h}^2}\bigg] \nonumber \\
=&\ \frac{1+8\alpha\Lambda}{8\pi G}\kappa_{\rm TH}.
\end{align}
Substituting this into (\ref{Swkeval}), we finally derive
\begin{align}
S_{\rm WK}=&\frac{1+8\alpha\Lambda}{4G}A_{\rm h}.
\end{align}
The above expression works also for the vacuum black hole represented by the metric function (\ref{sol-4dim-4}) in the singular theory with $1+8\alpha\Lambda =0$.

\subsection{ Case $\beta \ne 0$}
\label{app:entropy2}
The metric function of the new solution for $\beta\ne 0$ is given by Eq.~(\ref{sol4dim-1}), namely
\begin{align}
f(v,r)=k-\frac{2m(v)}{r}-\frac{\Lambda}{3}r^2.
\end{align}
The areal radius of the future outer trapping horizon ${r=r_{\rm h}}$ is determined by
\begin{align}
k-\frac{2m(v)}{r_{\rm h}}-\frac{\Lambda}{3}r_{\rm h}^2=0 \label{TH-relation1}
\end{align}
and its surface gravity is given by
\begin{align}
\kappa_{\rm TH}=\frac{m(v)}{r_{\rm h}^2}-\frac{\Lambda}{3}r_{\rm h}=\frac{k-\Lambda r_{\rm h}^2}{2r_{\rm h}},\label{kappa-TH}
\end{align}
where we used Eq.~(\ref{TH-relation1}) at the last equality.

With the metric function (\ref{sol4dim-1}), $Q^{vr}$ reduces to
\begin{align}
Q^{vr}= \frac{[1+2\Lambda(4\alpha+\beta)](3m-\Lambda r^3)}{24\pi G\, r^2}
-\frac{\beta\,{\dot m}}{2\pi G\, r^3},
\end{align}
of which value on the trapping horizon is
\begin{align}
Q^{vr}|_{r=r_{\rm h}}=&\ \frac{[1+2\Lambda(4\alpha+\beta)](k-\Lambda r_{\rm h}^2)}{16\pi G r_{\rm h}}-\frac{\beta\,{\dot m}}{2\pi G r_{\rm h}^3} \nonumber \\
=&\ \frac{[1+2\Lambda(4\alpha+\beta)]\kappa_{\rm TH}}{8\pi G}-\frac{\beta\,{\dot m}}{2\pi G r_{\rm h}^3},
\end{align}
where we used Eqs.~(\ref{TH-relation1}) and (\ref{kappa-TH}).
Finally, the Wald-Kodama dynamical entropy is computed to give
\begin{align}
S_{\rm WK}=&\Big[1+2\Lambda(4\alpha+\beta)\Big]\frac{A_{\rm h}}{4G}-\frac{\beta V^{(k)}_2{\dot m}}{G\kappa_{\rm TH} r_{\rm h}} \nonumber \\
=&\Big[1+2\Lambda(4\alpha+\beta)\Big]\frac{A_{\rm h}}{4G}-\frac{2\beta V^{(k)}_2{\dot m}}{G(k-\Lambda r_{\rm h}^2)},\label{WKentropy-app}
\end{align}
where we used Eqs.~(\ref{Swkeval}) and~(\ref{kappa-TH}).

In the static case, where $m$ is constant, Eq.~(\ref{WKentropy-app}) reduces to the formula for the Wald entropy given in~\cite{criticalgravity}.
In critical gravity, in particular, the Wald entropy is vanishing because ${1+2\Lambda(4\alpha+\beta)=0}$ holds.
It was shown that the Wald entropy is vanishing in critical gravity for any stationary black hole which is an Einstein spacetime~\cite{aor2017}.



\begin{references}
\bibitem{Weinberg:1974tw}
 S.~Weinberg,
``Problems in Gauge Field Theories''.
In the proceedings of the XVII International Conference on High Energy Physics, editor J.~R.~Smith (Rutherford Laboratory, Chilton, Didcot, Oxfordshire), III-59.
\bibitem{Deser:1975nv}
  S.~Deser,
  ``The State of Quantum Gravity,''
  Conf.\ Proc.\ {\bf C750926} (1975) 229.
In the proceedings of the conference on Gauge Theories and Modern Field Theory, editors R.~Arnowitt and P.~Nath (MIT press, Cambridge, Massachusetts).
\bibitem{Stelle:1976gc}
  K.~S.~Stelle,
  Phys.\ Rev.\ D {\bf 16}, 953  (1977).
\bibitem{Salvio2018}
A.~Salvio,
e-Print: arXiv:1804.09944 [hep-th].
\bibitem{Zwiebach:1985uq}
  B.~Zwiebach,
  Phys.\ Lett.\ {\bf B156}, 315 (1985).
\bibitem{bb1996}
M. C.~Bento and O.~Bertolami,
Phys. Lett. {\bf B368}, 198 (1996).
\bibitem{bht2009}
E. A.~Bergshoeff, O.~Hohm, and P. K.~Townsend,
Phys. Rev. Lett. {\bf 102}, 201301 (2009).
e-Print: arXiv:0901.1766 [hep-th]
\bibitem{bht2009b}
E. A.~Bergshoeff, O.~Hohm, and P. K.~Townsend,
Phys. Rev. D {\bf 79}, 124042 (2009).
e-Print: arXiv:0905.1259 [hep-th]
\bibitem{no2009}
M.~Nakasone and I.~Oda,
Prog. Theor. Phys. {\bf 121}, 1389 (2009).
\bibitem{ohta2012}
N.~Ohta,
Class. Quant. Grav. {\bf 29}, 015002 (2012).
\bibitem{ls2009}
Y.~Liu and Y.-w.~Sun,
JHEP {\bf 0904}, 106 (2009).
e-Print: arXiv:0903.0536 [hep-th]
\bibitem{oda2009}
I.~Oda,
JHEP {\bf 0905}, 064 (2009).
e-Print: arXiv:0904.2833 [hep-th]
\bibitem{mo2012}
K.~Muneyuki and N.~Ohta,
Phys. Rev. D {\bf 85}, 101501 (2012).
\bibitem{criticalgravity}
H.~Lu and C. N.~Pope,
Phys. Rev. Lett. {\bf 106}, 181302 (2011).
\bibitem{criticalgravity-higher}
S.~Deser, H.~Liu, H.~Lu, C. N.~Pope, T. C.~Sisman, and B.~Tekin,
Phys. Rev. D {\bf 83}, 061502 (2011).
\bibitem{wald}
R. M.~Wald, {\it General Relativity}, (University of Chicago Press,
1984).
\bibitem{hayward1994}
S.~A.~Hayward,
Phys. Rev. D {\bf 49}, 6467 (1994).
\bibitem{nm2008}
M.~Nozawa and H.~Maeda,
Class. Quant. Grav. {\bf 25}, 055009 (2008).
\bibitem{chargedVaidya-GR}
S.~Chatterjee, S.~Ganguli, and A.~Virmani,
Gen. Rel. Grav. {\bf 48}, 91 (2016).


\bibitem{po2006}
J.~Podolsk\'y and M.~Ortaggio,
Class. Quantum Grav. {\bf 23}, 5785 (2006).

\bibitem{opz2008}
M.~Ortaggio, J.~Podolsk\'y, and M.~\v{Z}ofka,
Class. Quantum Grav. {\bf 25}, 025006 (2008).

\bibitem{f1974}
V.~P.~Frolov,
Zh.~Eksp.~Teor.~Fiz. {\bf 66}, 813 (1974); {\it Sov. Phys. JETP}, {\bf 39}, 393.

\bibitem{chargedVaidya-EGB}
A. E.~Dominguez and E.~Gallo,
Phys. Rev. D {\bf 73}, 064018 (2006).
\bibitem{f(R)}
T. P.~Sotiriou and V.~Faraoni,
Rev. Mod. Phys. {\bf 82}, 451 (2010).
\bibitem{maeda2010}
H.~Maeda,
Phys. Rev. D {\bf 81}, 124007 (2010).
\bibitem{maeda2011}
H.~Maeda,
JHEP {\bf 1102}, 039 (2011).
\bibitem{fs2013}
M.~Flory and I.~Sachs,
Phys. Rev. D {\bf 88}, 044034 (2013).
\bibitem{flory-thesis}
M.~Flory,
Master's Thesis, 2013. (Available in \url{http://www.theorie.physik.uni-muenchen.de/TMP/theses/mflory_thesis.pdf}.)
\bibitem{tm2005}
T.~Torii and H.~Maeda,
Phys. Rev. D {\bf 71} 124002, (2005);
T.~Torii and H.~Maeda,
Phys. Rev. D {\bf 72}, 064007 (2005).
\bibitem{kodama1980}
H. Kodama,
Prog. Theor. Phys. {\bf 63}, 1217 (1980).
\bibitem{wald1993}
R. M.~Wald,
Phys. Rev. D {\bf 48}, 3427 (1993).
\bibitem{iyerwald1994}
V. Iyer and R. M. Wald,
Phys. Rev. D {\bf 50}, 846 (1994).
\bibitem{jkm1994}
T.~Jacobson, G.~Kang, and R. C.~Myers,
Phys. Rev. D {\bf 49}, 6587 (1994).
\bibitem{pad}
T.~Padmanabhan, {\it Gravitation: Foundations and Frontiers}, (Cambridge University Press, 2010).
\bibitem{lovelock}
D. Lovelock,
J. Math. Phys. {\bf 12}, 498 (1971).
\bibitem{hma1999}
S. A.~Hayward, S.~Mukohyama, and M. C.~Ashworth,
Phys. Lett. {\bf A256}, 347 (1999).
\bibitem{aor2017}
G.~Anastasiou, R.~Olea, and D.~Rivera-Betancour,
e-Print: arXiv:1707.00341 [hep-th].


\end{references}
\end{document}